\documentclass[twocolumn,aps,superscriptaddress,showpacs]{revtex4}
\usepackage{amssymb}
\usepackage{amsmath}
\usepackage{graphicx}
\usepackage[normalem]{ulem}
\usepackage[dvips]{color}

\begin{document}

\title{Tensor force induced isospin-dependence of short-range nucleon-nucleon correlation and high-density behavior of nuclear symmetry energy}

\author{Chang Xu}
\affiliation{Department of Physics and Astronomy, Texas A$\&$M
University-Commerce, Commerce, Texas 75429-3011,
USA}\affiliation{School of Physics, Nanjing University, Nanjing
210008, China}
\author{Bao-An Li\footnote{Corresponding author, Bao-An\_Li$@$Tamu-Commerce.edu}}
\affiliation{Department of Physics and Astronomy, Texas A$\&$M
University-Commerce, Commerce, Texas 75429-3011, USA}

\begin{abstract}
Quantitative information on the tensor force induced
isospin-dependence of short-range nucleon-nucleon correlation
(SRC) extracted from recent J-Lab experiments was used in
constraining the high-momentum tails of single nucleon momentum
distributions in both symmetric nuclear matter (SNM) and pure
neutron matter (PNM). Its effects on the Equations of State of SNM
and PNM as well as the nuclear symmetry energy are investigated.
It is found that the tensor force induced isospin-dependence of
SRC softens significantly the nuclear symmetry energy especially
at supra-saturation densities.
\end{abstract}


\maketitle

\section{Introduction}
Recently, much efforts have been devoted to study the
isospin-dependence of short-range nucleon-nucleon correlations
(SRC) in nuclei both experimentally \cite{Tan03,pia,sub,Bag10} and
theoretically \cite{alv,sch,nef,Dic04,Frick05,Sar05}. For reviews,
see, e.g., Refs.\ \cite{fra,Arr11}. It is particularly exciting to
note that experiments at the Jefferson Lab (J-Lab) have shown that
about 20\% of nucleons in $^{12}$C are correlated. Most
interestingly, the strength of $np$ SRC is about 20 times that of
$pp~(nn)$ SRC \cite{sub}. Theoretical studies have revealed that
the dominance of $np$ over $pp~(nn)$ SRC is a direct consequence
of the tensor force acting in the $np$ deuteron-like state
\cite{pia,alv,sch,nef}. Moreover, it was found that
\cite{sch,Sar05} the isospin-dependence of SRC is robust and does
not depend on the exact parameterization of the nucleon-nucleon
force, the type of nucleus, or the exact ground-state wave
function used to describe the nucleons \cite{sub}. Because the
local density of SRC pairs in nuclei is estimated to reach that
expected in the core of neutron stars, it has been repeatedly
speculated that the observed isospin-dependence of the SRC may
have significant effects on the Equation of State (EOS) of cold
dense neutron-rich nucleonic matter and thus properties of neutron
stars \cite{pia,sub}. Nevertheless, a quantitative evaluation of
the effects has not been carried out yet.

Nuclear symmetry energy $E_{sym}(\rho)$, which encodes the energy
related to neutron-proton asymmetry in the nuclear matter EOS, is
a vital ingredient in the theoretical description of neutron stars
and of the structure of neutron rich nuclei and reactions
involving them. Since the density-dependence of $E_{sym}(\rho)$ is
still the most uncertain part of the EOS of neutron-rich nucleonic
matter especially at supra-saturation densities, to better
determine the $E_{sym}(\rho)$ has become a major goal of both
nuclear physics and astrophysics
\cite{ireview98,ibook01,dan,bar,li1,Sum94,Lat04,Ste05a,Xuli10a,Xuli10b,LWC05,tsa,Cen09,xia}.
The $E_{sym}(\rho)$ can be approximated as the difference between
the energy per nucleon in pure neutron matter (PNM) and symmetric
nuclear matter (SNM), i.e.,
$E_{sym}(\rho)=E_{PNM}(\rho)-E_{SNM}(\rho)$. The
isospin-dependence of SRC affects differently the EOSs of PNM and
SNM at high densities, it is thus expected to play an important
role in determining the high density behavior of $E_{sym}(\rho)$.
In fact, it has long been known that the tensor force acting in
isosinglet neutron-proton pairs influences the high density
behavior of $E_{sym}(\rho)$ \cite{Pan72,Wir88a,bro1,xuli,Lee1}.
However, the strength of the in-medium tensor force and its
effects on the SRC were not so clear. The newly available and more
quantitative information on the tensor force induced
isospin-dependence of SRC may allow us to better understand
effects of the tensor force on the EOS and symmetry energy of
dense neutron-rich nucleonic matter. In this work, within a
phenomenological model for nuclear matter we show that the tensor
force induced isospin-dependence of SRC stiffens significantly the
EOS of SNM especially at super-saturation densities. However, it
has almost no effect on the EOS of PNM. Consequently, the tensor
force acting in neutron-proton isosinglet pairs significantly
lowers the symmetry energy of dense neutron-rich nucleonic matter.

\section{Nucleon momentum distribution including effects of short-range nucleon-nucleon correlations}
To treat properly SRC effects has been a challenging task for
nuclear many-body theories. For the present study, the most
relevant question is how the SRC affects the one-body nucleon
momentum distribution $n(k)$. In fact, this question has been
studied both experimentally and theoretically for a long time.
Fortunately, many interesting results have already been well
established and can be used reliably in our present work. For a
comprehensive review, we refer the reader to the book by Antonov
et al. \cite{ant}. The $n(k)$ contains information not only about
the nuclear mean-field but also the SRC. For a Fermi gas of
uncorrelated nucleons at zero temperature, the $n(k)$ is simply a
step function of one for $k\leq k_F$, and zero for $k>k_F$ where
$k_F$ is the nucleon Fermi momentum. The SRC, however, will
deplete states below the Fermi surface and make the states above
it partially occupied, leading to a high-momentum tail in $n(k)$.
This feature has been well established in various microscopic and
phenomenological models \cite{ant}. Thus, in a nucleus, the
momentum of a low-momentum nucleon with $k\leq k_F$ is balanced by
the rest of the nucleus; however, a high-momentum nucleon with
$k\geq k_F$ is almost always balanced by only one other nucleon
and the two nucleons form a correlated pair \cite{Bag10}. It is
also well known that the SRC can be caused by either the repulsive
core of the central force or the tensor part of the
nucleon-nucleon interaction. The experimental finding that the
$np$ SRC dominates over the $nn~(pp)$ one indicates that the
tensor force instead of the repulsive core is mainly responsible
for the high-momentum tail of $n(k)$.

\begin{figure}[htb]
\centering
\includegraphics[width=8.7cm]{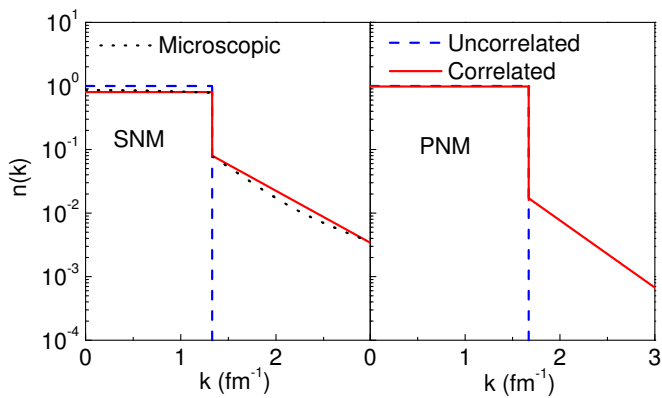}
\caption{(Color online) Single nucleon momentum distribution in
symmetric nuclear matter and pure neutron matter.} \label{n(k)}
\end{figure}
To investigate how the strength of tensor force affects the
high-momentum tail of $n(k)$, some phenomenological methods have
been shown to be particularly useful \cite{ant}. Instead of
seeking for exact many-body wave functions containing the SRC as
in microscopic many-body theories, the phenomenological methods
include the SRC in a physically very transparent way and allow one
to easily examine effects of the tensor force on the momentum
distribution $n(k)$. For instance, Dellagiacoma \textit{et. al}
have derived formulas for $n(k)$ explicitly including the tensor
force induced SRC for finite nuclei \cite{dell,tra}. It was shown
that the probability of finding nucleons with high momentum
increases with the strength of tensor force (or equivalently the
percentage of D-wave mixture) \cite{dell,tra}. With the same
D-wave percentage, it was also found that the high-momentum tails
of $n(k)$ for different nuclei are very close to each other
\cite{dell,tra}. This finding is consistent with the conclusion of
microscopic calculations that the behavior of $n(k)$ at high
momenta ($k>2$ fm$^{-1}$) is almost independent of the mass number
\cite{ant,fan}. Moreover, the high momentum behavior of $n(k)$ for
symmetric nuclear matter was shown to be very similar to those of
finite nuclei \cite{ant,fan,cio}. Results obtained using the
y-scaling analysis of inclusive electron scattering data for
$^{2}$H,  $^{3,4}$He, $^{12}$C, $^{56}$Fe also support this
conclusion \cite{cio2}.

Here we parameterize the $n(k)$ of nucleons in both SNM and PNM as
\begin{eqnarray}\label{n(k)}
n(k) & = & a  \,\,\, (k \leq k_F) \\ \nonumber
     & = & e^{b\,k} \,\,\, (k > k_F),
\end{eqnarray}
under the normalization condition \cite{fan} that
\begin{equation}
\frac{3}{k_F^3}\int_0^{\infty} n(k) k^2dk =1.
\end{equation}
The $a$ and $b$ are parameters determined using the experimental
findings from the J-Lab experiments. For SNM, we take $80\%
~(20\%)$ nucleons as uncorrelated (correlated) as indicated by the
J-Lab experiments and model calculations as we discussed earlier.
Thus, for SNM we have
\begin{eqnarray}
\frac{3}{k_F^3}\int_0^{k_F} n(k) k^2dk &=&0.8,\\\nonumber
\frac{3}{k_F^3}\int_{k_F}^{\infty} n(k) k^2dk&=&0.2.
\end{eqnarray}
The required parameter $a=0.8$ and $b$ is obtained numerically
from the integration $\frac{3}{k_F^3}\int_{k_F}^{\infty} e^{b\,k}
k^2dk =0.2$.  The value of $b$ depends on the Fermi momentum and
thus the density. At nuclear matter saturation density, $b=
-1.899$. Shown in the left panel of Fig.\ref{n(k)} is the $n(k)$
of symmetric nuclear matter at saturation density. The blue dashed
line denotes the $n(k)$ in an ideal Fermi gas. The red solid line
is our parameterization. The black dotted line is the
parameterization given by Ciofi degli Atti et.al. \cite{cio},
which is a fit to the result of variational many-body calculations
\cite{fan}. It is clearly seen that our parameterization is very
close to the microscopic single nucleon momentum distribution. For
PNM, the SRC is induced only by the repulsive core. Utilizing the
experimental finding that only 2\% of nucleons in $^{12}$C can
form the $nn$- or $pp$- type SRC pairs, we have for PNM
\begin{eqnarray} \frac{3}{2k_F^3} \int_0^{2^{\frac{1}{3}}k_F} n(k)
k^2dk &=&0.98,\\\nonumber
\frac{3}{2k_F^3}\int_{2^{\frac{1}{3}}k_F}^{\infty} n(k) k^2dk
&=&0.02.
\end{eqnarray}
The resulting parameter $a$ is 0.98. Again, the parameter $b$ is
density-dependent and has to be found numerically. At nuclear
matter saturation density, $b= -2.435$. The corresponding $n(k)$
for PNM is shown in the right panel of Fig.\ref{n(k)}. As
mentioned above, the big difference in $n(k)$ for SNM and PNM is
caused by the tensor-force induced SRC in SNM.

\section{Effects of short-range nucleon-nucleon correlations on the EOS and symmetry energy of dense neutron-rich nucleonic matter}
We now examine effects of the isospin-dependence of SRC on the EOS
and symmetry energy of dense neutron-rich nucleonic matter by
using the single nucleon momentum distributions in SNM and PNM
constrained by the J-Lab experiments within a phenomenological
approach. With momentum-dependent nuclear interactions, not only
the kinetic but also the potential energy are affected by the
high-momentum tail due to the SRC. Let us first examine separately
effects of the SRC on the kinetic and potential energy per nucleon
in both SNM and PNM. While the average kinetic energy per nucleon
in a Fermi gas of independent nucleons is simply
$E_{kin}=\frac{3}{5} \frac{\hbar^2 k_F^2}{2m}$, with correlated
nucleons it is given by
\begin{eqnarray}
E_{kin} = \alpha\int_0^{\infty} \frac{\hbar^2 k^2}{2m} n(k) k^2 dk,
\end{eqnarray}
where $\alpha=\frac{3}{k_F^3}$ and $\frac{3}{2k_F^3}$ for SNM and
PNM, respectively.
\begin{figure}[htb]
\centering
\includegraphics[width=8cm]{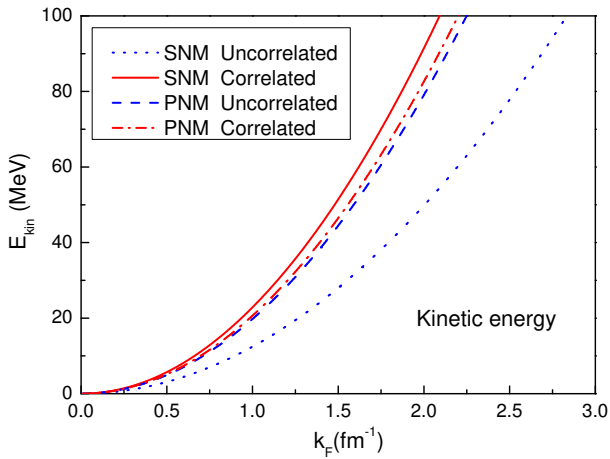}
\caption{(Color online) Average kinetic energy per nucleon
$E_{kin}$ for both symmetric nuclear matter and pure neutron
matter.} \label{E_{kin}}
\end{figure}
We compare in Fig.\ref{E_{kin}} the average kinetic energies for
both SNM and PNM calculated with and without the SRC. As one
expects, the SRC increases the $E_{kin}$ significantly for SNM but
only little for PNM because of the strong isospin-dependence of
the SRC. More quantitatively, for SNM at saturation density, the
$E_{kin}$ with the SRC ($E_{kin}(k_F^0)\simeq40$ MeV) is about
twice of that ($E_{kin}(k_F^0)\simeq22$ MeV) for the Fermi gas of
uncorrelated nucleons. However, for PNM, the increase in $E_{kin}$
due to the SRC induced only by the repulsive core is very small.
The different changes in $E_{kin}$ for SNM and PNM due to the
isospin-dependence of the SRC has a dramatic effect on the kinetic
part of the nuclear symmetry energy. In the Fermi gas without the
SRC, the kinetic energy contribution to the symmetry energy is
$E_{sym}^{kin}=E_{PNM}^{kin}-E_{SNM}^{kin}=(2^{\frac{2}{3}}-1)
(\frac{3}{5} \frac{\hbar^2 k_F^2}{2m})$, which is always positive
with increasing density. However, because the SRC induced increase
in $E_{kin}$ is much larger for SNM than PNM, the kinetic
contribution to the symmetry energy now becomes negative. Thus,
the kinetic part of $E_{sym}$ and its density-dependence are
strongly affected by the tensor-force induced SRC.
\begin{figure}[htb]
\centering
\includegraphics[width=8cm]{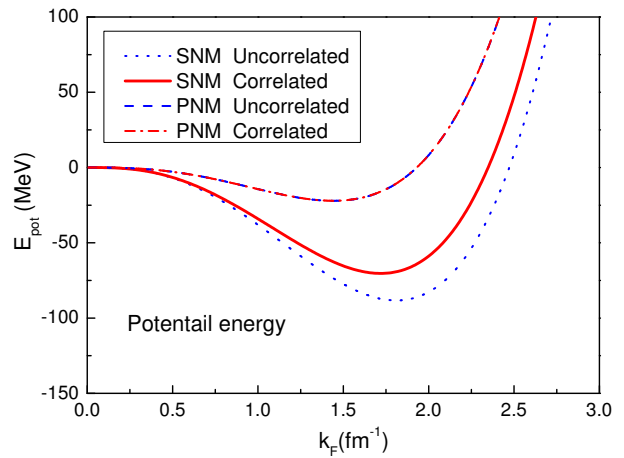}
\caption{(Color online) Average potential energy per nucleon
$E_{pot}$ for both symmetric nuclear matter and pure neutron
matter.} \label{E_{pot}}
\end{figure}

The average potential energy per nucleon $E_{pot}$ can be
calculated using potential energy density functionals. For
extracting information about the EOS of dense nuclear matter from
heavy-ion reactions, several momentum-dependent single-particle
potentials, such as the GBD (Gale-Bertsch-Das Gupta) \cite{gal},
BGBD (Bombaci-Gale-Beretsch-Das Gupta) \cite{bom} and MDI
(Momentum-Dependent Interaction)\cite{Das03} phenomenological
potentials, have been widely used. In calculating the
corresponding EOSs with these potentials only the step function
for the momentum distribution of a Fermi gas was used. Thus,
effects of the SRC were not considered. Here we use the MDI
interaction and the nucleon momentum distributions including the
SRC effects. The MDI potential energy per nucleon can be written
as \cite{Das03}
\begin{eqnarray}
E_{pot} & = & \frac{A}{2} \frac{\rho}{\rho_0} + \frac{B}{\sigma+1}
\frac{\rho^{\sigma}}{\rho_0^{\sigma}} \\ \nonumber & + & \frac{C}{\rho\rho_0}
\int_0^{\infty} \int_0^{\infty}
\frac{n(k_1)n(k_2)}{1+(\vec{k_1}-\vec{k_2})^2/\Lambda^2}d\vec{k_1}d\vec{k_2},
\end{eqnarray}
where $\sigma=4/3$ and the parameter $\Lambda=1.0 k_F^0$
\cite{Das03}. The $n(k_1)$ and $n(k_2)$ are the one-body momentum
distribution of nucleon-1 and nucleon-2. For SNM, as usual, the
parameters $A=-114.079$ MeV, $B=107.154$ MeV and $C=-0.0274$ MeV
are obtained from fitting its three empirical saturation
properties, namely the vanishing pressure, the average binding
energy of $E=-16$ MeV, and the incompressibility of $K_0=220$ MeV
at the saturation density $\rho_0=0.16fm^{-3}$. For PNM, we obtain
the three parameters $A=-120.570$ MeV, $B=102.080$ MeV and
$C=-7.61\times 10^{-4}$ MeV by satisfying the well established
theoretical constrains on the low density PNM EOS given in
Refs.\cite{pie10,car}, namely, $E=1.7$ MeV at $k_F=0.4$ fm$^{-1}$
and $E=4.2$ MeV at $k_F=0.8$ fm$^{-1}$, as well as the symmetry
energy $E_{sym}=31$ MeV at $\rho_0$. Shown in Fig.\ref{E_{pot}} is
a comparison of $E_{pot}$ with and without the SRC for both SNM
and PNM. Similar to the case of $E_{kin}$, the SRC increases the
magnitude of $E_{pot}$ significantly for SNM but has negligible
effect for PNM. Consequently, the tensor force induced SRC also
softens the potential part of the symmetry energy
$E_{sym}^{pot}=E_{PNM}^{pot}-E_{SNM}^{pot}$.
\begin{figure}[htb]
\centering
\includegraphics[width=8cm]{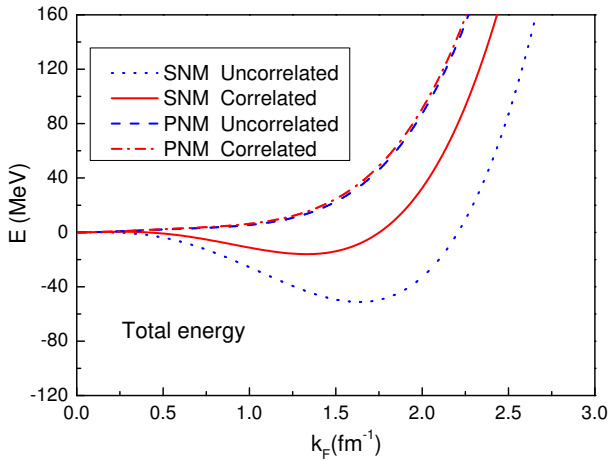}
\caption{(Color online) Total binding energy per nucleon for both
symmetric nuclear matter and pure neutron matter.} \label{E_{tot}}
\end{figure}
\begin{figure}[htb]
\centering
\includegraphics[width=8cm]{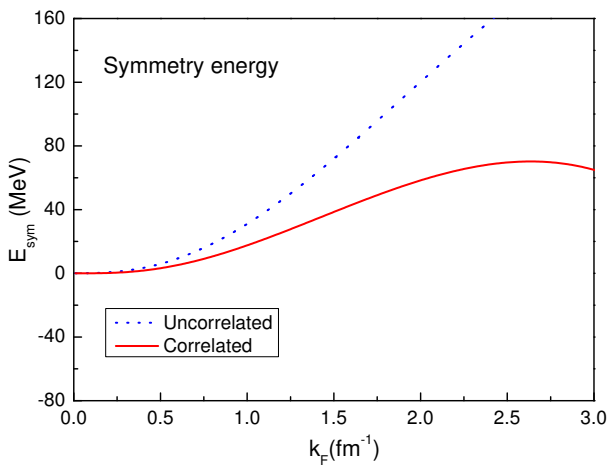}
\caption{(Color online) Symmetry energy with (correlated) and
without (uncorrelated) the shore-range nucleon-nucleon
correlations.} \label{E_{sym}}
\end{figure}
Shown in Fig.\ref{E_{tot}} are the EOSs of SNM and PNM with and
without the SRC. The corresponding symmetry energies are presented
in Fig.\ref{E_{sym}}. It is interesting to see that the tensor
force induced SRC has a significant impact on the symmetry energy
especially at supra-saturation densities. The symmetry energy
obtained in calculations with the SRC is much softer especially
for $k_F>2.0$ fm$^{-1}$. Microscopic many-body theories have shown
clearly that the isosinglet nucleon-nucleon interaction dominates
the symmetry energy \cite{Bom91,Die03}. It was also known that the
tensor force acting in the isosinglet channel affects
significantly the high-density behavior of nuclear symmetry energy
\cite{Pan72,Wir88a,bro1,xuli,Lee1}. The finding here that the SRC
softens the symmetry energy is thus not surprising qualitatively.
In fact, the short-range repulsive tensor force may even lead to
negative symmetry energies at supra-saturation densities as first
predicted by Pandharipande and Garde in 1972 within a variational
many-body approach \cite{Pan72}. However, we never had before the
kind of quantitative information on the isospin-dependence of the
SRC due to the tensor force and the repulsive core from
experiments. The results presented here thus represent a
significant progress in our understanding about the tensor force
effects on the EOS and symmetry energy of dense neutron-rich
nucleonic matter.

\section{Summary}
In summary, the quantitative information on the tensor force
induced isospin-dependence of SRC from the recent J-Lab
experiments was used in constraining the high-momentum tails of
single nucleon momentum distributions in SNM and PNM. The latter
are then used in evaluating the EOSs of SNM and PNM as well as the
symmetry energy of neutron-rich nucleonic matter. It is found that
the isospin-dependence of SRC stiffens significantly the EOS of
SNM but has little effect on that of PNM. Consequently, the
nuclear symmetry energy especially at supra-saturation densities
is significantly softened. This finding confirms the critical role
of the tensor force acting in neutron-proton isosinglet pairs in
determining the high-density behavior of nuclear symmetry energy
predicted by various microscopic and phenomenological nuclear
many-body theories.

\section{Acknowledgement}
We would like to thank Lie-Wen Chen for helpful discussions. This
work is supported by the US National Science Foundation grant
PHY-0757839, the Texas Coordinating Board of Higher Education
grant No.003565-0004-2007, the National Aeronautics and Space
Administration under grant NNX11AC41G issued through the Science
Mission Directorate, the National Natural Science Foundation of
China (Grants 10735010, 10775068, and 10805026) and by the
Research Fund of Doctoral Point (RFDP), No. 20070284016.

\end{document}